\documentclass[11pt,a4paper]{article}
\usepackage{amsmath,amssymb,bm,epsfig,color,graphicx,cite,braket,mathrsfs,slashed,multirow,amsfonts,latexsym}

\renewcommand{\thefootnote}{\fnsymbol{footnote}}

\usepackage{hyperref,xcolor}
\hypersetup{
    colorlinks=true,
    linktocpage=true,
    linkcolor=blue,
    filecolor=blue,      
    urlcolor=blue,
    citecolor=blue,
    }
    
\usepackage[compat=1.1.0]{tikz-feynhand}

\usepackage{siunitx}
\begin{document}

\title{
\begin{flushright}
\begin{minipage}{0.2\linewidth}
\normalsize
CTPU-PTC-23-10 \\*[50pt]
\end{minipage}
\end{flushright}
{\Large \bf Updated Constraints and Future Prospects \\ on Majoron Dark Matter\\*[20pt]}}

\author{
Kensuke Akita$^{1}$\footnote{
\href{mailto:kensuke8a1@ibs.re.kr}{kensuke8a1@ibs.re.kr}}\ \ \ \ and\ \ \ \
Michiru Niibo$^{2,3}$\footnote{
\href{mailto:g2370609@edu.cc.ocha.ac.jp}{g2370609@edu.cc.ocha.ac.jp}}\\*[20pt]
$^1${\it \small
Particle Theory and Cosmology Group, Center for Theoretical Physics of
the Universe, }\\{\it\small Institute for Basic Science (IBS), Daejeon, 34126, Korea} \\
$^2${\it \small
Department of Physics, Graduate School of Humanities and Sciences,}\\{\it\small  Ochanomizu University, Tokyo 112-8610, Japan} 
\\
$^3${\it\small
Cosmology, Gravity, and Astroparticle Physics Group, Center for
Theoretical Physics of the Universe, }\\{\it\small  Institute for Basic Science (IBS), Daejeon,
34126, Korea}
\\*[50pt]
}

\date{
\centerline{\small \bf Abstract}
\begin{minipage}{0.9\linewidth}
\medskip \medskip \small 
Majorons are (pseudo-)Nambu-Goldstone bosons associated with lepton number symmetry breaking due to the Majorana mass term of neutrinos introduced in the seesaw mechanism. They are good dark matter candidates since their lifetime is suppressed by the lepton number breaking scale.  We update constraints and discuss future prospects on majoron dark matter in the singlet majoron models based on neutrino, gamma-ray, and cosmic-ray telescopes in the mass region of MeV--10 TeV.
\end{minipage}
}

\maketitle{}
\thispagestyle{empty}
\addtocounter{page}{-1}
\clearpage
\noindent\hrule
\tableofcontents
\noindent\hrulefill

\renewcommand{\thefootnote}{\arabic{footnote}}
\setcounter{footnote}{0}
%\vspace{35pt}

%%%%%%%%%%%%%%%%%%%%%%%%%%%%%%%%%%%%%%%%%%%%%%%%%%%%%%%%%%%%%%%%%%%%%%%%%%%%%%%%%%%%%%%%%%%%%%%%%%
%%%%%%%%%%%%%%%%%%%%%%%%%%%%%%%%%%%%%%%%%%%%%%%%%%%%%%%%%%%%%%%%%%%%%%%%%%%%%%%%%%%%%%%%%%%%%%%%%%
%%%%%%%%%%%%%%%%%%%%%%%%%%%%%%%%%%%%%%%%%%%%%%%%%%%%%%%%%%%%%%%%%%%%%%%%%%%%%%%%%%%%%%%%%%%%%%%%%%
%%%%%%%%%%%%%%%%%%%%%%%%%%%%%%%%%%%%%%%%%%%%%%%%%%%%%%%%%%%%%%%%%%%%%%%%%%%%%%%%%%%%%%%%%%%%%%%%%%
\section{Introduction}
\label{sec1}

The non-zero mass of left-handed neutrinos revealed by oscillation experiments is nothing more than a clear deviation from the Standard Model (SM).
The most promising mechanism to explain the origin of the small neutrino masses would be the seesaw mechanism \cite{Minkowski:1977sc, Mohapatra:1979ia, Gell-Mann:1979vob, Yanagida:1979as,Schechter:1980gr}. In this model, the neutrino mass is naturally suppressed from the electroweak scale by the Majorana mass of right-handed neutrinos. In the seesaw mechanism, lepton number symmetry is broken due to the Majorana mass term. If the lepton number symmetry is actually a global symmetry as in the SM and becomes spontaneously broken at low energies, there is an associated Nambu-Goldstone (NG) boson, called a majoron \cite{Chikashige:1980qk,Chikashige:1980ui,Gelmini:1980re,Schechter:1981cv}. Since the majoron can be a long-lived particle due to its interactions suppressed by the lepton symmetry breaking scale, it can be an attractive candidate for dark matter (DM) \cite{Rothstein:1992rh, Berezinsky:1993fm, Lattanzi:2007ux,Bazzocchi:2008fh,Gu:2010ys,Frigerio:2011in, Lattanzi:2013uza, Queiroz:2014yna,Dudas:2014bca,Wang:2016vfj,Garcia-Cely:2017oco,Abe:2020dut,Manna:2022gwn} if the majoron mass is generated from an explicit (soft) symmetry breaking term originating from gravitational effects \cite{Rothstein:1992rh,Akhmedov:1992hi} or other mechanisms \cite{Mohapatra:1982tc,Langacker:1986rj,Ballesteros:2016euj,Ballesteros:2016xej,Frigerio:2011in,Gu:2010ys}.

X- and gamma-ray and cosmic-ray experiments have led the way in DM searches by exploring DM decays and annihilations into photons and charged particles (see e.g.\ Refs.\ \cite{Ibarra:2013cra, PerezdelosHeros:2020qyt} for a review), and majoron DM is no exception. Since a majoron decays into photons and charged particles at loop level, 
X- and gamma-ray and cosmic-ray observations put 
constraints on majoron DM in the keV-GeV DM mass range\footnote{Particle decay experiments also constrain the majoron model via majoron production \cite{Garcia-Cely:2017oco,Heeck:2019guh}. For the majoron mass of $\gtrsim {\rm MeV}$, these constraints are weaker than cosmological observations.} \cite{Bazzocchi:2008fh,Lattanzi:2013uza,Garcia-Cely:2017oco}. 
Another complementary approach to investigate majoron DM is detecting neutrino signatures. 
In particular, the two-body decays into neutrinos at tree level produce a monochromatic neutrino signal, which can be easily distinguished from background.
Even though neutrinos are the most difficult particles to detect in SM, the neutrino channel offers a complementary information in studying majoron DM. 
The two-body decay rates of majorons into neutrinos at tree level depend only on the symmetry breaking scale and the majoron mass, while the two-body decay rates into photons and charged particles at loop level additionally depend on a coupling constant such as a Yukawa coupling in the seesaw mechanism. 
Neutrino telescopes also constrain majoron DM in the MeV-GeV mass range \cite{Garcia-Cely:2017oco}.

Despite the difficulties of neutrino detection, neutrino telescopes with very large volumes, such as Super-Kamiokande \cite{Olivares-DelCampo:2017feq,Palomares-Ruiz:2007egs,Frankiewicz:2016nyr} and IceCube \cite{IceCube:2017rdn,IceCube:2021kuw,IceCube:2023ies} are currently improving the constraints on the DM annihilation and lifetime (see e.g.\ Refs.\ \cite{Arguelles:2019ouk,Arguelles:2022nbl} and references theirin).
Furthermore, the next-generation large neutrino detectors such as Hyper-Kamiokande \cite{Bell:2020rkw}, JUNO \cite{Akita:2022lit}, DUNE \cite{Cuesta:2022mut}, IceCube-Gen2 \cite{Kochocki:2023lhh}, KM3NET \cite{Sanguineti:2023qfa} and P-ONE \cite{P-ONE:2020ljt} are expected to significantly improve the sensitivity on the neutrino signals from DM in the MeV-PeV mass range, and it is important to update the restrictions on majoron DM in light of the progress of these neutrino telescopes. 
In addition, the next-generation gamma-ray telescope such as CTA \cite{Pierre:2014tra} might also improve the other visible signals from majoron DM with the mass of TeV.

In this paper, we update the work by Garcia-Cely and Heeck \cite{Garcia-Cely:2017oco} on the constraints on majoron DM in the singlet majoron model from neutrino line signatures and other visible signatures by majoron decays.
The production mechanism of majoron DM with mass of ${\rm keV}\text{--}10\ {\rm TeV}$ are proposed \cite{Rothstein:1992rh,Berezinsky:1993fm,Frigerio:2011in,Abe:2020dut,Manna:2022gwn} and majoron DM can lead to observable neutrino signatures for energies from MeV to $10\,{\rm TeV}$. We update the constraints on majoron DM using the latest observational data and extend the constraints in the mass region of majoron from up to $100\ {\rm GeV}$ to $10\ {\rm TeV}$.
We also discuss future prospects on the sensitivity of majoron DM from the next-generation neutrino and gamma-ray telescopes.

This paper is organized as follows: in section \ref{sec:model}, we briefly summarize the singlet majoron model and neutrino spectrum from majoron DM we consider in this work. In section \ref{sec:constraint}, we review the relevant neutrino experiments and present our updated constraints and future sensitivity of majoron DM. Section \ref{sec:conclusions} is devoted to conclusions. In appendix~\ref{app:comparison}, we compare our results to those in Ref.\ \cite{Garcia-Cely:2017oco} and explain the updated constraints, providing references to the data and prior analyses used in our study.

\section{Singlet majoron model}
\label{sec:model}
In this section, we briefly review the singlet majoron model and neutrino signals from majoron dark matter we consider, following Ref.~\cite{Garcia-Cely:2017oco,Heeck:2019guh}.
We assume the majoron, $J$, is massive, a pseudo-Nambu-Goldstone boson, and dark matter.
The production mechanisms of majoron DM with mass of ${\rm keV}\text{--}10\ {\rm TeV}$ has been proposed in Ref.~\cite{Rothstein:1992rh,Berezinsky:1993fm,Frigerio:2011in,Abe:2020dut,Manna:2022gwn}.
The signals from the majoron decays discussed below are independent of the mechanism of the majoron mass generation and the majoron production.

\subsection{Lagrangian and the majoron decay rates}

The fact that the neutrino mass is below eV scale can be naturally explained by the seesaw mechanism with heavy right-handed Majorana masses $M_{R}\gg v \sim 10^{2}\,{\rm GeV}$, where $v$ is the higgs vacuum expectation value (vev). This Majorana mass $M_{R}$ can be generated by the spontaneous lepton number symmetry breaking, U(1)$_{\rm L}$ (or U(1)$_{\rm B-L}$), and the associated Nambu-Goldstone (NG) boson is called a majoron \cite{Chikashige:1980qk,Chikashige:1980ui,Gelmini:1980re,Schechter:1981cv}. The interaction Lagrangian in the singlet majoron model is described as follows:
\begin{align}
    \mathcal{L} = -\lambda_{D} \Phi^{*}\overline{E_{L}}\nu_{R}- \frac{\lambda_{R}}{2}\overline{\nu_{R}^{c}}\Sigma\nu_{R}+{\rm h.c.},
    \label{Lsingletmajoron}
\end{align}
where $\Phi$ and $E_{l}$ are the higgs doublet and lepton doublet in the SM, respectively, $\Sigma$ is a newly introduced complex scalar singlet with the lepton number of $L(\Sigma)=-2$ and $\nu_{R}$ is a lepton singlet, which is a right-handed neutrino. 
$\lambda_{D}$ and $\lambda_{R}$ are coupling constants. 
We assume right-handed neutrinos have three species. The different number of right-handed neutrino species will not change the following discussions.
The whole Lagrangian conserves U(1)$_{\rm L}$. After the spontaneous U(1)$_{\rm L}$ breaking at the scale of $f$, the new scaler obtains the vev, $\Sigma(x)=(f+\sigma(x) + iJ(x))/\sqrt{2}$, which introduces the Majorana mass for the right-handed neutrinos, $M_{R} = \lambda_{R}f/\sqrt{2}$. Here, $J$ is the majoron. Below the energy scale of electroweak symmetry breaking, the SM higgs also obtains the vev $v$ with $\Phi(x)=(0,v+h(x))^{T}/\sqrt{2}$, which introduces the Dirac mass for mixing between left- and right-handed neutrinos $m_{D} = \lambda_{D}v/\sqrt{2}$. The whole mass matrix for neutrinos is diagnalized as
\begin{align}
    \begin{pmatrix}
        0&m_{D}\\m_{D}^{T}&M_{R}
    \end{pmatrix}
    =V^{*}{\rm diag}(m_{1}\cdots m_{6})V^{\dagger},
\end{align}
 with a unitary matrix $V$, and neutrinos in the mass basis are determined as $n=n_{R} + n_{R}^{c}$, where,
\begin{align}
    n_{R} = V^{\dagger}
    \begin{pmatrix}
        \nu_{L}^{c}&N_{R}
    \end{pmatrix}.
\end{align}
The neutrino couplings to the majoron, $J$, $Z$-boson and $W$-boson are given in the mass basis as
\begin{align}
    \mathcal{L}_{J}=&-\frac{iJ}{2f}\sum_{i,j=1}^6\overline{n}_{i}\left[\gamma_{5}(m_{i}+m_{j})\left(\frac{1}{2}\delta_{ij}-{\rm Re}C_{ij}\right)+ i(m_{i}-m_{j}){\rm Im}C_{ij}\right]n_{j}, \label{Jnucoupling} \\
    \mathcal{L}_{Z}=&-\frac{g_w}{4\cos\theta_w}\sum_{i,j=1}^6\overline{n}_{i}\slashed{Z}\left[i{\rm Im} C_{ij}-\gamma_5{\rm Re} C_{ij} \right]n_{j}, \\
    \mathcal{L}_{W}=&-\frac{g_w}{2\sqrt{2}}\sum_{i,j=1}^6\overline{l}_{i}B_{ik}\slashed{W}^-(1-\gamma^5)n_{j}+{\rm h.c.},
\end{align}
where
\begin{align}
    C_{ij} = \sum_{k=1}^{3}V_{ki}V^\ast_{kj},\ \ \ \ B_{ij}=\sum_{k=1}^3U_{ik}^lV_{kj}^\ast,
\end{align}
with a unitary matrix $U^l$ for the diagonalization of the charged lepton mass matrix which can be the identity matrix without loss of generality, $U_{ik}^l=\delta_{ik}$.

The seesaw mechanism works when $M_{R}\gg m_{D}$, so that we simply assume $f\gg v$. 
We further assume that a majoron has the mass with $m_{1},m_{2},m_{3}\ll m_{J}\ll m_{4},m_{5},m_{6}$. 
Then the majoron can decay into the two light left-handed neutrinos at tree level, $J\rightarrow \nu_{i}\nu_{i}\,(i=1,2,3)$.
In the seesaw limit, $M_{R}\gg m_{D}$, the decay rate into two neutrinos is,
\begin{align}
    \Gamma(J\rightarrow 2\nu) &\simeq \frac{m_{J}}{16\pi f^{2}}\sum_{i=1}^{3}m_{i}^{2}\\
    &\sim \frac{1}{3\times 10^{19}\,{\rm sec}}\left(\frac{m_{J}}{1\,{\rm MeV}}\right)\left(\frac{10^{9}\,{\rm GeV}}{f}\right)^{2}\left(\frac{\sum m_{i}^{2}}{10^{-3}{\rm eV^{2}}}\right), \label{eq:total decay rate into neutrinos}
\end{align}
so that the majoron has essentially long lifetime to be a dark matter candidate. We should note that for $m_J \gtrsim 10\ {\rm TeV}, J\rightarrow \nu\nu h (h)$ is a dominant decay channel compared to $J\rightarrow \nu\nu$ \cite{Dudas:2014bca}.
The majoron-neutrino-higgs coupling, $J\nu_i\nu_i(m_i/f)(1+h/v)^2$, is induced from Eq.~(\ref{Jnucoupling}).

\begin{figure}
    \centering
    \includegraphics[width=15cm]{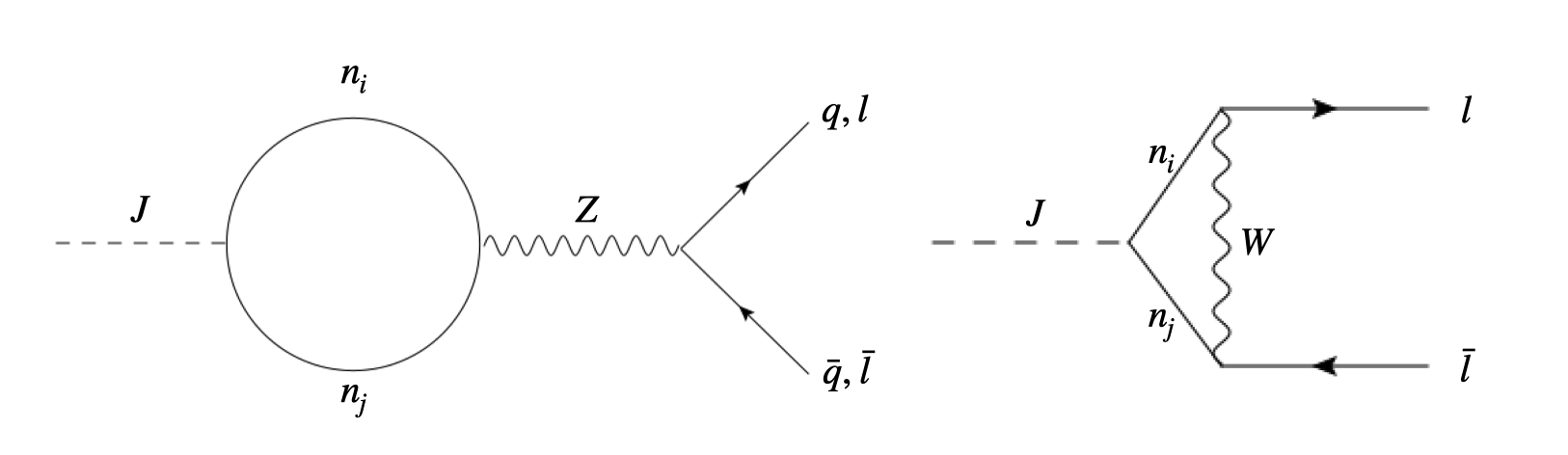}
    \caption{1-loop Feynman diagram for $J\rightarrow q\bar{q}, \,l\bar{l}$.}
    \label{fig:Jqq}
\end{figure}

\begin{figure}
    \centering
    \includegraphics[width=17cm]{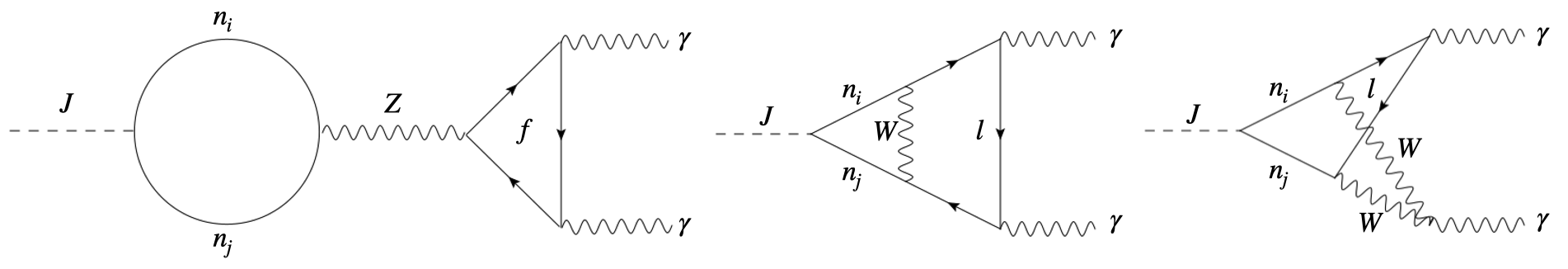}
    \caption{2-loop Feynman diagram for $J\rightarrow \gamma\gamma$.}
    \label{fig:Jgammagamma1}
\end{figure}

Majorons can also decay into two charged fermions, $J\rightarrow \bar{f}f$, and two photons, $J\rightarrow \gamma\gamma$, at one-loop and two-loop level, respectively. 
The Feynman diagrams for the two-body decays into quarks and charged leptons are shown in Figure.\ \ref{fig:Jqq}.
The decay rates into quarks and charged fermions in the seesaw limit are given by,
\begin{align}
    \Gamma(J\rightarrow q\bar{q})
    &\simeq\frac{3m_{J}}{8\pi}\left|\frac{m_{q}}{8\pi^{2}v}T_{3}^{q}{\rm tr}K\right|^{2},\\
    \Gamma(J\rightarrow l\bar{l'})
    &\simeq\frac{m_{J}}{8\pi}\left(\left|\frac{m_{l}+m_{l'}}{16\pi^{2}v}(\delta_{ll'}T_{3}^{l}{\rm tr}K+K_{ll'})\right|^{2}+\left|\frac{m_{l}-m_{l'}}{16\pi^{2}v}K_{ll'}\right|^{2}\right),
\end{align}
with $T_{3}^{d,l}=-1/2=-T_{3}^{u}$, and
\begin{align}
    K\equiv \frac{m_{D}m_{D}^{\dagger}}{vf}.\label{Eq:K}
\end{align}
The Feynman diagram of the two-body decay channel into photons is shown in Figure.\ \ref{fig:Jgammagamma1}, and the decay rate is \cite{Heeck:2019guh},
\begin{align}
\Gamma(J\rightarrow 2\gamma)
    &\simeq\frac{\alpha^{2}}{4096\pi^{7}}\frac{m_{J}^{3}}{v^{2}} |K'|^2,\label{eq:decay rate into photons}
\end{align}
with
\begin{align}
K'={\rm tr}K\sum_{f} N_{c}^{f}T_{3}^{f}Q_{f}^{2}h\left(\frac{m_{J}^{2}}{4m_{f}^{2}}\right)+\sum_l K_{ll}h\left(\frac{m_{J}^{2}}{4m_{l}^{2}}\right),\label{eq:Kprime}
\end{align}
where $\alpha$ is the fine-structure constant, $N_{c}^{q}=3,\, N_{c}^{l}=1$ are the factors accounting the number of colors, $Q_{c}^{f}$ is the electric charge, and 
\begin{align}
    h(x) &=-\frac{1}{4x}\left({\rm log}(1-2x+2\sqrt{x(x-1)})\right)^{2}-1.
\end{align}
Note that similar diagrams with the W-boson triangle loop to the left panel of Figure.~\ref{fig:Jgammagamma1} cancel with the diagrams with Faddeev-Popov ghosts \cite{Garcia-Cely:2016hsk}.

The cosmic microwave background (CMB), X- and gamma-ray, and cosmic-ray observations constrain DM decays into charged fermions and photons. In this model, those constraints are interpreted as the limit on the components of $K$, so that the constraints from these channels are always a combination of the lepton number symmetry breaking energy scale $f$ and the Yukawa coupling constant $\lambda_{D}$. On the other hand, the decay channels to neutrinos provide pure constraints on $f$.
Thus, the constraints on the majoron DM from $J\rightarrow \nu\nu$ and $J\rightarrow \bar{f}f$ are independent.
We can estimate $K\sim \lambda_D^2v/f$ and then $\Gamma(J\rightarrow 2\nu)\sim m_J\sum_im_i^2/f^2$ and $\Gamma(J\rightarrow \bar{f}f)\sim m_Jm_f^2\lambda_D^4/f^2$.
Thus, for $\lambda_D\sim 1$, the decay rates into charged leptons are larger than those into neutrinos.

The matrix $K$ can be rewritten by the seesaw parameters, using the Casas-Ibarra parametrization \cite{Casas:2001sr}, in the seesaw limit,
\begin{align}
K=\frac{1}{vf}U\sqrt{d_l}R^Td_hR^\ast\sqrt{d_l}U^\dag,
\label{Eq:K2}
\end{align}
where $d_l={\rm diag}(m_1,m_2,m_3) \ll d_h={\rm diag}(m_4,m_5,m_6)$ and $U$ 
is the Pontecorvo–Maki–Nakagawa –Sakata (PMNS) matrix.
$R$ is a complex orthogonal $3\times 3$ matrix, satisfying $m_D=iU\sqrt{d_l}R^T\sqrt{d_h}$.
This means that in principle, using low-energy neutrino parameters ($d_l$ and $U$) and the majoron parameters ($K$ and $f$), the seesaw mechanism Eq.\ (\ref{Lsingletmajoron}) can be reconstructed.

Finally, we comment on the other possible channels of the majoron decays. In the following, we neglect these channels. In general, this neglect would be a conservative choice since the total signal from the majoron decays is reduced. However, a signal from a majoron decay may sometimes mimic background in another signal, weakening the constraints on the majoron DM.
We leave a careful analysis of these neglected channels as future work.

First, note that the decays of majoron with mass range of our interest into light quarks, $J\rightarrow \bar{u}u$, $\bar{d}d$, $\bar{s}s$, $\bar{c}c$, should be appropriately replaced by decays into hadron. However, the majoron decay rate into hadrons is not well understood. In the following, we consider only $J\rightarrow \bar{b}b,\,\bar{t}t$ as the decay of quarks as in Ref.~\cite{Garcia-Cely:2017oco}.
For $m_J \gtrsim 200\ {\rm MeV}$ and $ 100\ {\rm GeV}$, the additional decay channels including gluons and $Z,W,h$-bosons such as $J\rightarrow gg,\ WW,\ ZZ,\ hh$ can be induced, respectively \cite{Heeck:2019guh}, but we neglect these channels for simplicity. 
At one-loop level, the decay rate is suppressed by the right-handed neutrino mass squared, $M_R^2$, since $\nu_L\text{--}\nu_R$ mixing is necessary to close the loop.
At two-loop level, the decay rates might be suppressed by $\alpha$ or the QCD coupling constant $\alpha_S$ at least compared to $J\rightarrow \bar{f}f$.
We should also note that even at one-loop level, photons are emitted by $J\rightarrow \bar{f}f\gamma$. This decay rate may be further suppressed by $\alpha$ or the right-handed neutrino masses~\cite{Garcia-Cely:2017oco}, compared with $J\rightarrow \bar{f}f$. We do not take into account this channel to constrain the majoron model. Depending on the production mechanism of majoron DM, the additional coupling between majoron and higgs is sometimes assumed (e.g., Refs.~\cite{Frigerio:2011in,Abe:2020dut}). We also neglect this effect on the majoron decay rates.

\subsection{Neutrino flux from majoron decays}

In Eq.\ \eqref{eq:total decay rate into neutrinos}, $\nu_{i}$ is a neutrino in the mass basis. Ignoring matter effects, the propagation of neutrino mass-eigenstates from the location of dark matter decay to the earth will not suffer neutrino oscillations, while neutrino telescopes detect flavor-neutrinos via the charged leptons produced by weak-interactions. The possibility that a majoron-induced neutrino $\nu_{i}$ is detected as a $\nu_{\beta}$ ($\beta = e,\mu,\tau$) is,
\begin{align}
    P(i\rightarrow \beta)
    = |\braket{\nu_{i}|\nu_{\beta}}|^{2}= |{U_{\beta i}}|^{2},
\end{align}
where $U_{\beta i}$ is the ($\beta,i$)-component of PMNS matrix.
Then the branching ratio of the majoron decay into flavor neutrinos is
\begin{align}
    \alpha_{\beta} = \frac{\sum_{i=1}^{3}|U_{\beta i}|^{2}m_{i}^{2}}{\sum_{i=1}^{3}m_{i}^{2}}.
\end{align}
The $\nu_{\beta}$ flux from the two-body decay of majoron DM in the Milky Way is,
\begin{align}
    \frac{d\Phi_{\nu_{\beta}}}{dE_{\nu}}
    =\frac{d\Phi_{\bar{\nu}_{\beta}}}{dE_{\nu}}
    =& \frac{\mathcal{D}}{4\pi m_{J}}\alpha_{\beta}\frac{\Gamma(J\rightarrow 2\nu)}{2}\frac{dN}{dE_\nu}\label{eq:neutrino flux},
\end{align}
where $dN/dE_\nu=2\delta(E_\nu-m_J/2)$ is the neutrino spectrum from the two-body decay of a majoron. 
We neglect the extragalactic $\nu_\beta$ flux for simplicity and a conservative purpose.
Since $\Gamma$ is the total decay rate including the decay into both neutrinos and anti-neutrinos, a factor of $1/2$ is necessary.
The astrophysical factor $\mathcal{D}$, which determines neutrino intensity from Milky Way, is defined as
\begin{align}
    \mathcal{D}=\int_{0}^{2\pi}dl\int^{\pi/2}_{-\pi/2}db \cos{b} \int_0^{s_{\rm max}} ds\ \rho\left(\sqrt{R_{\rm sc}^2-2sR_{\rm sc}\cos\psi+s^2} \right),
    \label{Dfactor}
\end{align}
where $\cos{\psi} = \cos{b}\cos{l}$, $R_{\rm sc}\simeq 8\ {\rm kpc}$, and $s_{\rm max} = \sqrt{(R_{\rm MW}^2 - \sin^{2}{\psi}R_{\rm sc}^{2})} + R_{\rm sc}\cos{\psi}$ with $R_{\rm MW}=40\ {\rm kpc}$.
The DM profile of the Milky Way galaxy is still unknown, which introduces uncertainty in determining the flux.
Navarro-Frenk-White (NFW) \cite{Navarro:1995iw}, Moore\cite{Moore:1999gc} and Isothermal\cite{1980ApJS...44...73B} correspond $(\mathcal{D}/10^{23}\,{\rm GeV cm^{-2}})=1.9,2.5,1.9$, respectively. Considering the fact that the square root of the $\mathcal{D}$-factor appears in the bound of $f$, the uncertainty in the DM profile does not significantly affect the analysis. In this paper, the NFW profile is considered.

The branching ratios $\alpha_{\beta}$ and the sum of squared light neutrino mass $\sum_{i=1}^{3}m_{i}^{2}$ depend on mass hierarchy. The mixing angles of PMNS matrix $U$ and the mass splitting $\Delta m_{12}$ and $|\Delta m_{23}|$ are measured \cite{Esteban:2020cvm}\footnote{\href{http://www.nu-fit.org}{NuFIT 5.2 (2022), www.nu-fit.org}}. We consider three different extreme hierarchies of light neutrinos displayed in Table.\ \ref{tab:hierarchy}. To figure out the most conservative limit, we use 3$\sigma$ lower bound of the mass splitting obtained from neutrino oscillation experiments \cite{Esteban:2020cvm} in normal and inverse hierarchy (NH, IH, respectively) cases.
We also use the most conservative upper bound obtained from CMB \cite{Planck:2018vyg} (PlanckTT+lowE) in the quasi-degenerated (QD) case. Since the decay rate of neutrino inducing channel depends on the sum of squared mass of neutrino linearly, we use the most conservative to figure out the extreme estimation limit in QD regime, though some certain combination of datasets provides much stronger constraint below $0.005\,{\rm eV}^{2}$. For the branching ratios, we use the best-fitted values \cite{Esteban:2020cvm}. Here we assume the massless lightest neutrinos for the NH and IH cases. Both the squared neutrino mass and the branching ratios used in this work are tabulated in Table.\ \ref{tab:hierarchy}.

\begin{table}[]
    \centering
    \begin{tabular}{cc|cccc}
        &&$\sum_{i=1}^{3}m_{i}^{2}$&$\alpha_{e}$&$\alpha_{\mu}$&$\alpha_{\tau}$\\ \hline
         Normal Hierarchy (NH)& $m_{1}\ll m_{2}\ll m_{3}$& $2.6\times 10^{-3}\,{\rm eV^{2}}$&$0.03$&$0.55$&$0.42$\\
         Inverse Hierarchy (IH)& $m_{3}\ll m_{1}\lesssim m_{2}$& $4.9\times 10^{-3}\,{\rm eV^{2}}$&$0.49$&$0.22$&$0.30$\\
         Quasi-Degenerate (QD)& $m_{1}\sim m_{2}\sim m_{3}$& $0.10\,{\rm eV^{2}}$&$1/3$&$1/3$&$1/3$
    \end{tabular}
    \caption{The definition of mass hierarchies and the sum of squared neutrino mass and the branching ratio in each mass hierarchy. We assume the lightest neturino is massless for the NH and IH cases. For the sum of squared neutrino mass, we use the lower bound \cite{Esteban:2020cvm} for NH and IH cases, and upper bound \cite{Planck:2018vyg} in QD case, while we use the best-fitted values \cite{Esteban:2020cvm} for the branching ratios $\alpha_{\beta}$.}
    \label{tab:hierarchy}
\end{table}

\section{Updated constraints and future prospects on Majoron Dark Matter}\label{sec:constraint}

\subsection{Result from neutrino signal}
\label{sec:result_nu}

We show the limits and future sensitivities on the lepton number breaking scale, $f$, from neutrino signals produced by majoron DM decays, using different analyses and the latest data of Borexino\cite{Borexino:2019wln}, KamLAND \cite{KamLAND:2021gvi}, Super-Kamiokande\cite{Super-Kamiokande:2011lwo,Super-Kamiokande:2021jaq,Palomares-Ruiz:2007egs,Frankiewicz:2016nyr,Olivares-DelCampo:2017feq}, IceCube \cite{IceCube:2021kuw,IceCube:2023ies}, and ANTARES \cite{Albert:2016emp, Arguelles:2022nbl}, and the expected setups of JUNO \cite{Akita:2022lit}, Hyper-Kamiokande \cite{Bell:2020rkw}, and P-ONE \cite{P-ONE:2020ljt,Arguelles:2022nbl}. We also show the cosmological constraint on the DM lifetime comes from CMB+BAO analysis as $\leq 250\,{\rm Gyr}$ \cite{Audren:2014bca,Enqvist:2019tsa,Nygaard:2020sow,Alvi:2022aam,Simon:2022ftd}.
While DUNE \cite{Arguelles:2022nbl} and KM3NeT \cite{Gozzini:2020dom,BasegmezDuPree:2021fpo,Miranda:2022kzs} will have the excellent potential to explore DM decays into neutrinos, the contributions of each flavor neutrinos to these searches are non-trivial. We leave analyses of DUNE and KM3NeT for majoron DM as future work.

\begin{figure}
    \centering
    \includegraphics[width=15cm]{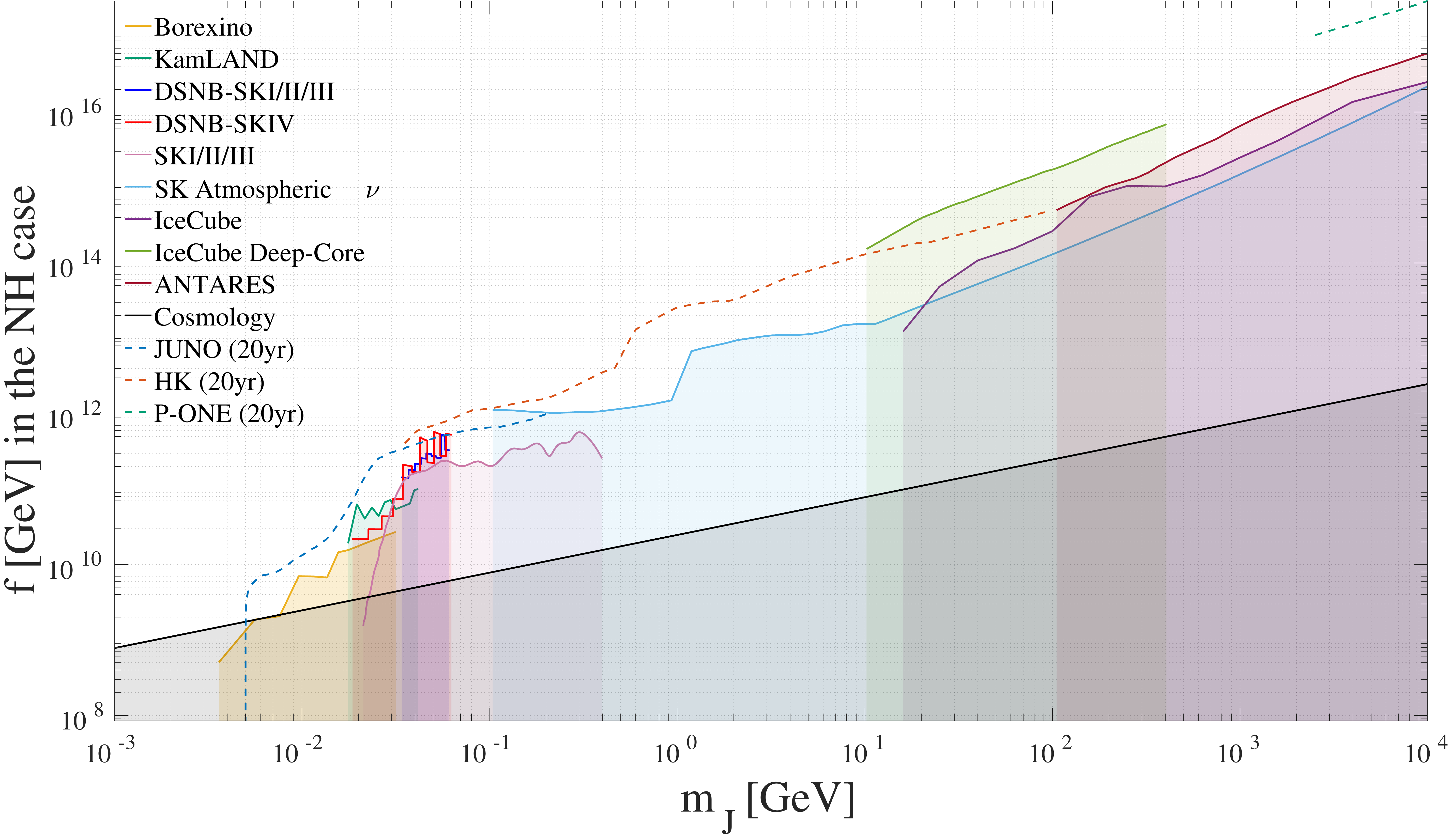}
    \caption{Lower bounds on the energy scale of the spontaneous lepton number symmetry breaking $f$ in NH case, as a function of the majoron DM mass. The black region corresponds to the cosmological constraint on the DM lifetime comes from CMB+BAO analysis as $\leq 250\,{\rm Gyr}$\cite{Audren:2014bca,Enqvist:2019tsa,Nygaard:2020sow,Alvi:2022aam,Simon:2022ftd}. The other colored regions with solid curves describe the current constraints from Borexino\cite{Borexino:2019wln} (yellow), KamLAND \cite{KamLAND:2021gvi} (green), Super-Kamiokande\cite{Super-Kamiokande:2011lwo,Super-Kamiokande:2021jaq,Palomares-Ruiz:2007egs,Frankiewicz:2016nyr,Olivares-DelCampo:2017feq} (red, blue, pink, light-blue), IceCube \cite{IceCube:2021kuw,IceCube:2023ies} (light-green, purple), and ANTARES \cite{Albert:2016emp, Arguelles:2022nbl}, and the dashed curves describe the expected sensitivities of future neutrino detectors, JUNO \cite{Akita:2022lit} (blue), HK\cite{Bell:2020rkw} (orange) and P-ONE \cite{P-ONE:2020ljt,Arguelles:2022nbl} (green).}
    \label{fig:figure NH}
\end{figure}

\begin{figure}
    \centering
    \includegraphics[width=15cm]{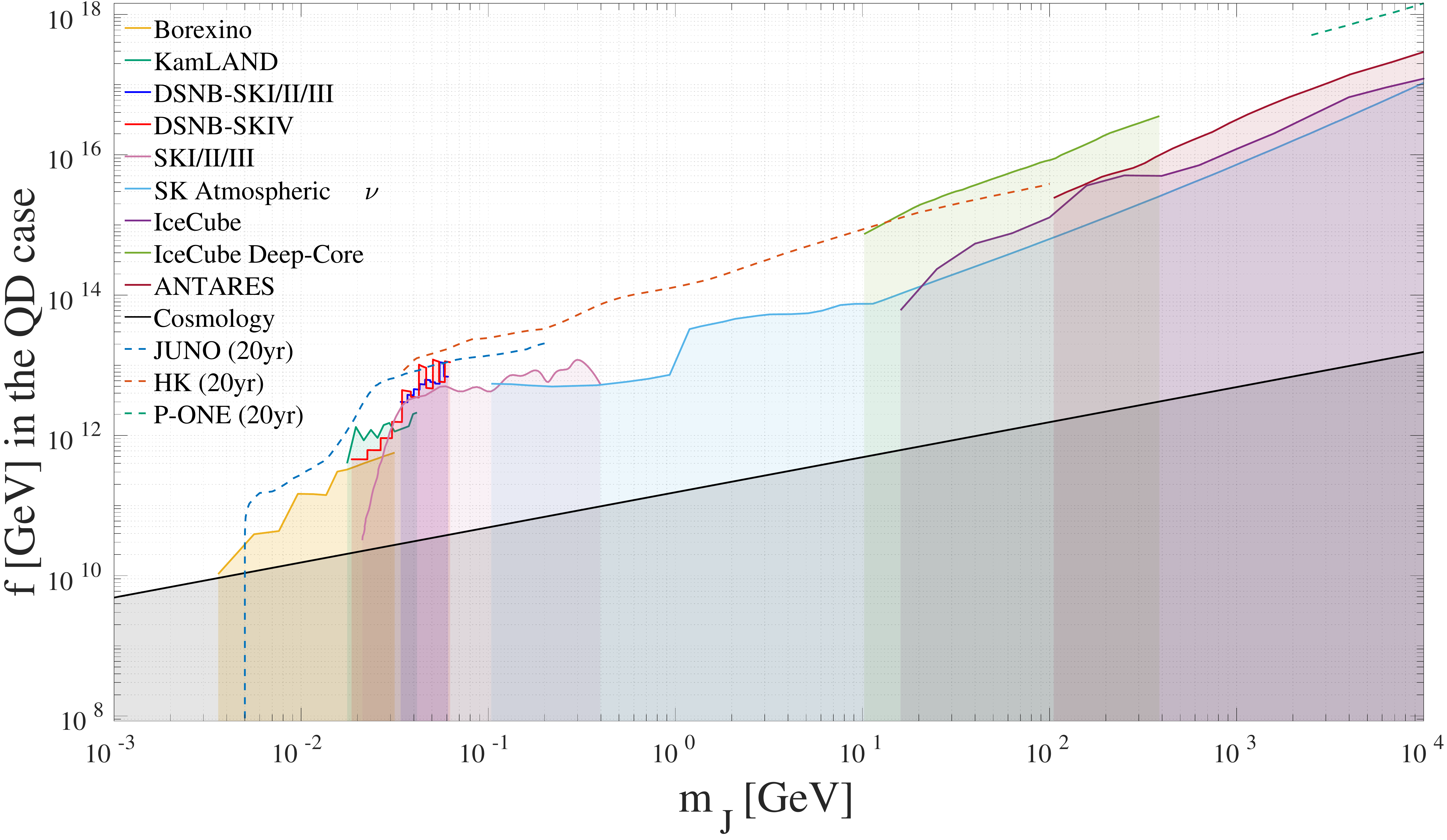}
    \caption{As in Figure.\ \ref{fig:figure NH} but for QD case.}
    \label{fig:figure QD}
\end{figure}

In Figures.\ \ref{fig:figure NH}, and \ref{fig:figure QD}, we show our constraints on neutrino lines from majoron DM in the NH, and QD cases, respectively. In the case of IH, the constraints appear in between the NH and QD cases. We do not show the results of the IH case. The constraints on $f$ in the QD case are the strongest, while those in the NH case are the weakest. This is mainly because $\Gamma(J\rightarrow 2\nu)\propto \sum_i m_i^2$ and in the NH case, $\nu_e$-flux is suppressed due to small $\alpha_e$.
The current constraints and expected sensitivities are shown with solid and dashed curves, respectively, and the cosmological constraint is shown in black in these figures.
As a whole, we can see that the limit tends to be stronger for larger masses. One reason for this result is that the decay rate is proportional to the DM mass and inversely proportional to $f$ squared.
Note that the constraints in three cases can be reproduced by rescaling the constraints in the case of NH, where the rescaling factors are $\sqrt{\alpha_{\beta}\sum_{i}m_{i}^{2}}$ for $\beta$-flavor neutrino detections. 
In the rest of this section, we briefly review those experiments and show our results.


\subsubsection*{Borexino}
Borexino is a neutrino detector using liquid scintillator designed for the spectral measurement of low–energy solar neutrinos in the Laboratori Nazionali del Gran Sasso in Italy. The Borexino collaboration has derived 90\% confidence level (C.L.) upper limits on the all-sky flux of $\bar{\nu}_{e}$ from unknown sources with the neutrino energy raging $1.8 \,{\rm MeV}\leq E_{\nu}\leq15.8\,{\rm MeV}$ during 2485 days of data-taking \cite{Borexino:2019wln}. We assume that the sources include DSNB and majoron DM and derive the upper limits on majoron DM induced neutrino flux $\Phi_{\bar{\nu}_{e}}$ (Eq.\ \eqref{eq:neutrino flux}) from 
\begin{align}
    \Phi_{\bar{\nu}_{e}}< \left(\frac{d\tilde{\Phi}}{dE}-\frac{d\Phi_{\rm DSNB}}{dE}\right)\Delta E \label{eq:constraint from DSNB},
\end{align}
where $\Delta E=1\,{\rm MeV}$ is the energy bin, $\tilde{\Phi}$ is the upper limits \cite{Borexino:2019wln} and $\Phi_{\rm DSNB}$ is the theoretical value of the neutrino flux from DSNB. We consider the same theoretical flux of DSNB discussed in \cite{Akita:2022lit}, assuming the progenitor star collapsing into $83\%$ neutron star and $17\%$ black hole \cite{Horiuchi:2017qja}. The mean neutrino spectrum from Fig.~6 of Ref.~\cite{Horiuchi:2017qja}. For definiteness, we consider only the normal hierarchy in the neutrino masses since the DSNB flux is almost the same both in the normal and inverted hierarchy (e.g., see  Ref.~\cite{Moller:2018kpn}). The result is shown in yellow in Figures.\ \ref{fig:figure NH}, and \ref{fig:figure QD}.

\subsubsection*{KamLAND}
KamLAND is a neutrino detector using 1 kton of liquid scintillator in Kamioka, Japan, starting data-taking in 2002. The KamLAND collaboration has derived the constraint at 90\% C.L.\ on $\bar{\nu}_{e}$ flux from DM self-annihilation with neutrino energy ranging $(8.3\text{--}30.8)\, {\rm MeV}$ during 4528.5 days of data-taking \cite{KamLAND:2021gvi}. We reinterpret the constraint on the annihilation cross section into the constraint on $f$, which is shown in green in Figures.\ \ref{fig:figure NH}, and \ref{fig:figure QD}.

\subsubsection*{Super-Kamiokande}
Super-Kamiokande (SK) is a water-Cherenkov detector with a fiducial volume of 22.5 ktons, started data-taking in 1996 in Kamioka, Japan. The data is divided into several phases, and we use the following four phases: SK-I ($T^{1}$ = 1497 days from 1996 to 2001), SK-II ($T^{2}$ = 794 days from 2002 to 2005), SK-III ($T^{3}$ = 562 days from 2006 to 2008) and SK-IV ($T^{4}$ = 2970 days from 2008 to 2018). The SK collaboration has derived 90\% C.L.\ upper limits on the all-sky flux of $\overline{\nu_{e}}$ from unknown sources with the neutrino energy raging $(17\text{--}30.5)\,{\rm MeV}$ \cite{Super-Kamiokande:2011lwo} in SK-I/II/III during 2853 days of data-taking, and $(9.3\text{--}31.3)\,{\rm MeV}$ in SK-IV during 2970 days of data-taking \cite{Super-Kamiokande:2011lwo}, respectively. The constraints on $f$ are obtained from Eq.\ \eqref{eq:constraint from DSNB} with $\Delta E=1.5\,{\rm MeV}$ for SK-I/II/III, and $\Delta E=2\,{\rm MeV}$ for SK-IV, respectively. In Figures.\ \ref{fig:figure NH}, and \ref{fig:figure QD}, the constraint from SK-I/II/III and SK-IV are shown in blue and red, respectively.

The SK collaboration also analyzes DSNB best fit and upper flux limit by performing an unbinned maximum likelihood fit \cite{Super-Kamiokande:2011lwo,Super-Kamiokande:2021jaq}, and the limitation on DM induced neutrinos in wider energy region can be obtained using the same methods. Ref.\ \cite{Olivares-DelCampo:2017feq} has already derived the constraint at 90\% C.L.\ on $\bar{\nu}_{e}$ flux from DM self-annihilation with the neutrino energy ranging $(5\text{--}200)\, {\rm MeV}$. We reinterpret the constraint on the annihilation cross section into the constraint on $f$, which is shown in pink in Figures.\ \ref{fig:figure NH}, and \ref{fig:figure QD}.

In addition to the data in the energy range $(16\text{--}88)\,{\rm MeV}$ to detect DSNB, the data at the higher energy range focused on atmospheric neutrinos also provides constraint on DM induced neutrino flux. Ref.\ \cite{Yuksel:2007ac} and \cite{Palomares-Ruiz:2007egs} derived the constraint on $\nu_{\mu}+\bar{\nu}_{\mu}$ flux from DM self-annihilation and decay, respectively, including measurements by the Frejus, SK (with less than 1500 days of data-taking) and AMANDA detectors \cite{Super-Kamiokande:2004pou,Frejus:1994brq,AMANDA:2002pgr,IceCube:2007jwc,IceCube:2005sgi,Super-Kamiokande:2005mbp,Gaisser:2002jj,Honda:2004yz,Super-Kamiokande:2007uxr}. We reinterpret the 90\% C.L.\ constraint on the lifetime of DM in the DM mass range $(0.1\text{--}100)\,{\rm GeV}$\cite{Palomares-Ruiz:2007egs}. The constraint on the DM lifetime in the DM mass range $(1\text{--}100)\,{\rm GeV}$ is updated in Ref.\ \cite{Frankiewicz:2016nyr} with SK data during 4223 days. The whole limits are shown in light-blue in Figures.\ \ref{fig:figure NH}, and \ref{fig:figure QD}. In Figure.\ \ref{fig:figure NH}, the constraint from atmospheric $\nu_{\mu}+\bar{\nu}_{\mu}$ flux shown in light-blue is stronger than one from DSNB $\nu_{e}+\bar{\nu}_{e}$ flux shown in pink. This is because the decay rate of majoron DM into neutrinos is propotinal to neutrino mass squared, and the first generation neutrino which electron-flavor neutrino mostly mix to is massless in NH case.

\subsubsection*{IceCube}

IceCube \cite{IceCube:2016zyt} is a Cherenkov-type detector using one cubic kilometer of ice underneath the South Pole. It has 78 vertical strings in a hexagonal grid with 60 digital optical modules and additional 8 vertical strings with more dense digital optical modules in the center of the detector. The whole picture of the detector is given in Figure.\ 1 in Ref.\ \cite{IceCube:2017rdn}, where the former and latter strings are shown as black and red dots, respectively. The central area surrounded by a blue line is called the DeepCore sub-detector.
The IceCube collaboration has  derived the constraint the 90\% C.L.\ constraints on the $\nu_{e}+\bar{\nu}_{e}$, $\nu_{\mu}+\bar{\nu}_{\mu}$, and $\nu_{\tau}+\bar{\nu}_{\tau}$ fluxes from DM decay in the Galactic Center with the DM mass ranging $(16\text{--}4\times 10^{4})\, {\rm GeV}$ using 5 years of data \cite{IceCube:2023ies}.
The IceCube collaboration has also derived the 90\% C.L.\ constraints on the full-sky fluxes of $\nu_{e}+\bar{\nu}_{e}$, $\nu_{\mu}+\bar{\nu}_{\mu}$, and $\nu_{\tau}+\bar{\nu}_{\tau}$ from DM self-annihilation with the neutrino energy ranging $(5\text{--}200)\, {\rm GeV}$ using 8 years of data measured by Icecube including DeepCore \cite{IceCube:2021kuw}.  
We reinterpret the constraints on the lifetime \cite{IceCube:2023ies} and the annihilation cross section \cite{IceCube:2021kuw} into the constraints on $f$, as shown in green and brown, respectively. IceCube collaboration has derived the constraints on all three flavors of DM-induced neutrinos in Ref.\ \cite{IceCube:2023ies, IceCube:2021kuw}, and we combine the three constraints on $f$ obtained from different three flavors.

\subsubsection*{ANTARES}
ANTARES is the neutrino telescope installed within the deep of Mediterranean Sea, running from 2008 to 2022 to search astrophysical high-energy neutrinos. It is now dismantled, and a next-generation neutrino detector, KM3NeT is under construction closely to the site to take over the aim \cite{Zegarelli:2023jzu}.
Ref.\ \cite{Arguelles:2022nbl} has derived the constraint on the DM lifetime,
based on the constraint on DM annihilation using $\mu\text{--}$neutrino data \cite{Albert:2016emp}. We reinterpret the constraint into $f$
as shown in brown in Figures.\ \ref{fig:figure NH}, and \ref{fig:figure QD}.

\subsubsection*{JUNO}
The Jiangmen Underground Neutrino Observatory (JUNO) \cite{JUNO:2015zny} is a 20 kton neutrino detector made of linear alkylbenzene liquid scintillator (${\rm C_{6}H_{5}C_{12}H_{25}}$) to be built in Jiangmen, China. We and collaborators have derived the 90\% C.L.\ expected constraint on the full-sky flux of $\nu_{e}+\overline{\nu_{e}}$ from DM self-annihilation and decay with the neutrino energy ranging $\mathcal{O}(10\,{\rm MeV})\leq E_{\nu} \leq 100\, {\rm MeV}$ in 20 years of data-taking \cite{Akita:2022lit}. We reinterpret the constraint on the lifetime into the constraint on $f$, which corresponds to dashed blue curves in Figures.\ \ref{fig:figure NH} and \ref{fig:figure QD}. Those figures indicate that JUNO can updated the current constraints within $(2\text{--}3)$-fold in very wide mass region $(5\text{--}200)\,{\rm MeV}$, thanks to its large volume and high energy resolution.

\subsubsection*{Hyper-Kamiokande}
\begin{figure}
    \centering
    \includegraphics[width=15cm]{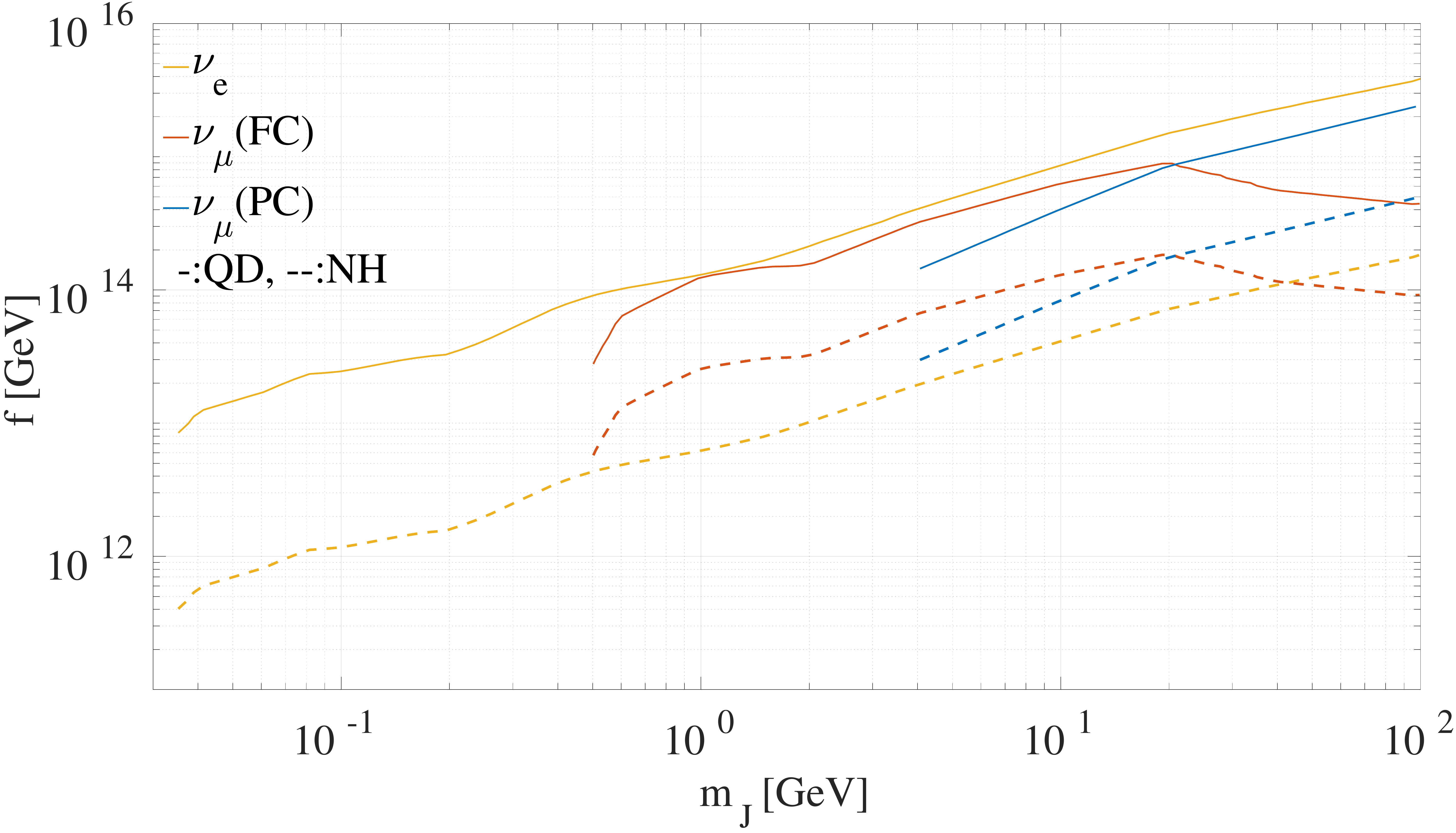}
    \caption{The reinterpreted constraints on $f$ from the HK experiment assuming 20 years of data-taking and 70\% of neutrino tagging efficiency \cite{Bell:2020rkw}. The yellow, red, and blue curves correspond to $\nu_{e}+\bar{\nu}_{e}$, $\nu_{\mu}+\bar{\nu}_{\mu}$(FC), and $\nu_{\mu}+\bar{\nu}_{\mu}$(PC) respectively, and dashed and solid curves correspond to the NH and QD cases. One can see that the $\nu_{e}+\bar{\nu}_{e}$ detection provides the strongest constraint in the QD case, while the combination of $\nu_{e}+\bar{\nu}_{e}$ ($m_{J}<500\,{\rm MeV}$), $\nu_{\mu}+\bar{\nu}_{\mu}$(FC) ($500\,{\rm MeV}<m_{J}<20\,{\rm GeV}$), and $\nu_{\mu}+\bar{\nu}_{\mu}$(PC) ($20\,{\rm GeV}<m_{J}$) provides the strongest constraint in the NH case. We omitted the IH case since the comparison of the constraints between the three channels is the same as the QD case.}
    \label{fig:hk}
\end{figure}

Hyper-Kamiokande (HK) \cite{Hyper-Kamiokande:2018ofw} is a water Cherenkov detector under constraction in Kamioka, Japan as a successor of SK. It is designed to have 187 kton of fiducial volume and to start data-taking in 2027. 
The expected constraint on the full-sky DM induced flux of $\nu_{e}+\bar{\nu}_{e}$ in the energy range $17\, {\rm MeV}\lesssim E_{\nu} \leq 50\, {\rm GeV}$ and $\nu_{\mu}+\bar{\nu}_{\mu}$ in the energy range $250\,{\rm MeV} \lesssim E_{\nu} \leq 50\, {\rm GeV}$ has been revealed in a recent work \cite{Bell:2020rkw}. The analysis classifies $\nu_{\mu}+\bar{\nu}_{\mu}$ fluxes into two categories. First, there are fully-contained (FC) events, in which all energy is deposited in the inner detector. On the other hand, there are partial-contained (PC) events, in which energetic muons leave the inner detector and deposit energy in the outer detector. 
The analysis provides the 90\% C.L.\ expected constraint on the full-sky DM induced flux of $\nu_{e}+\bar{\nu}_{e}$ in the energy range $17\, {\rm MeV}\leq E_{\nu} \leq 50\, {\rm GeV}$, $\nu_{\mu}+\bar{\nu}_{\mu}$ (FC) in the energy range $250\,{\rm MeV} \leq E_{\nu} \leq 50\, {\rm GeV}$ and $\nu_{\mu}+\bar{\nu}_{\mu}$ (PC) in the energy range $2\,{\rm GeV} \leq E_{\nu} \leq 50\, {\rm GeV}$, respectively, assuming 20 years of data-taking and 70\% of neutrino tagging efficiency \cite{Bell:2020rkw}. 
We reinterpret the constraint on the annihilation cross section into the constraints on $f$, which are shown in Figure.\ \ref{fig:hk}. In the figure, yellow, red, and blue curves correspond to $\nu_{e}+\overline{\nu_{e}}$, $\nu_{\mu}+\bar{\nu}_{\mu}$(FC), and $\nu_{\mu}+\bar{\nu}_{\mu}$(PC) respectively, and dashed and solid curves correspond to the NH and QD cases. We omitted the IH case in Figure.\ \ref{fig:hk} since the comparison of the strengths between the three channels does not change from the QD case. 
As the diagram indicates, the $\nu_{e}+\overline{\nu_{e}}$ detection provides the strongest constraint in QD case, while the combination of $\nu_{e}+\bar{\nu}_{e}$ ($m_{J}<500\,{\rm MeV}$), $\nu_{\mu}+\bar{\nu}_{\mu}$(FC) ($500\,{\rm MeV}<m_{J}<20\,{\rm GeV}$), and $\nu_{\mu}+\bar{\nu}_{\mu}$(PC) ($20\,{\rm GeV}<m_{J}$) provides strongest constraint in NH case. Due to the low energy cuts applied to $\mu\text{--}$ neutrinos signatures, $e\text{--}$ neutrinos signatures provide the constraint on the model for the DM mass below $0.5\,{\rm GeV}$, which gets significantly weaker in NH case. 
The strongest constraints from HK are shown as dashed orange curves in Figures.\ \ref{fig:figure NH} and \ref{fig:figure QD}.
Those figures indicate that HK will update the current constraint nearly one order of magnitude or even more in the mass region $(0.1\text{--}10)\,{\rm GeV}$ in the QD case. For the DM mass below $0.5\,{\rm GeV}$, the constraints only come from $e\text{--}$ neutrinos detection in our analysis, and therefore, do not update the current constraints drastically.

\subsubsection*{P-ONE}
The Pacific Ocean Neutrino Experiment (P-ONE) \cite{P-ONE:2020ljt} is a proposed cubic-kilometer scale neutrino telescope, planned to be installed within the Pacific Ocean underwater infrastructure of Ocean Networks Canada.
Ref.\ \cite{Arguelles:2022nbl} has derived the expected sensitivity on the DM lifetime, based on the $\mu\text{--}$neutrinos effective area for charged current interactions evaluated by the collaboration in Ref.\ \cite{P-ONE:2020ljt}. We reinterpret the expected sensitivity into $f$ within 20 years of data-taking, assuming that the difference of the DM profiles does not affect the constraint on $f$.
The results are shown as dashed green curves in Figures.\ \ref{fig:figure NH}, and \ref{fig:figure QD}.

\subsection{Result from other visible signals}
\label{sec:result_other}
\begin{figure}
    \centering
    \includegraphics[width=15cm]{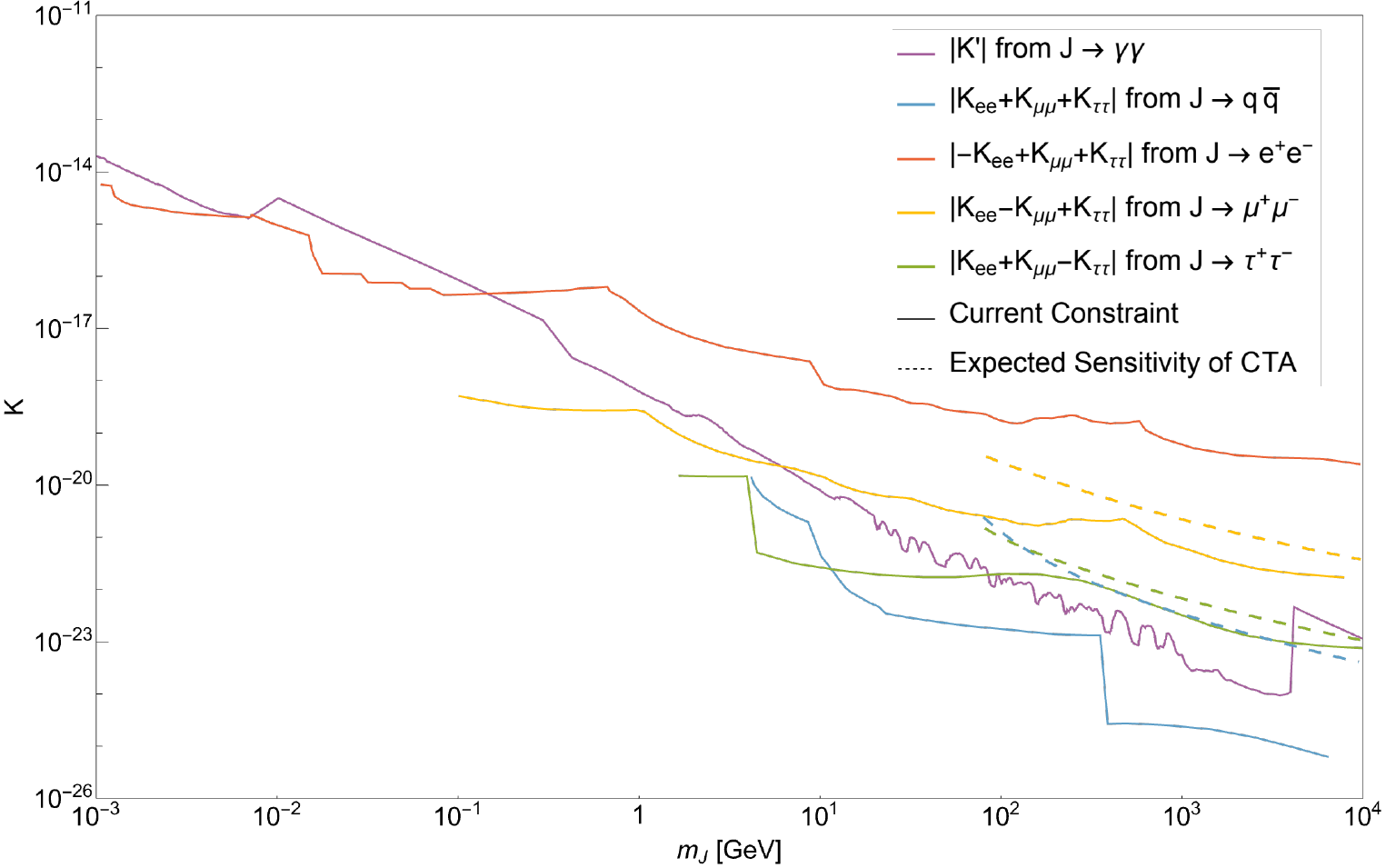} 
    \caption{Upper bounds on the linear combination of the components of $K$, defined in Eq.\ \eqref{Eq:K} and $|K'|$ defined in Eq.\ \eqref{eq:Kprime}. The solid purple curve shows the constraints from $J\rightarrow \gamma\gamma$ ($1\,{\rm MeV}\text{--}7\,{\rm MeV}$: INTEGRAL/SPI \cite{Fischer:2022pse}, $7\,{\rm MeV}\text{--}430\,{\rm MeV}$: 
    COMPTEL/EGRAT\cite{Yuksel:2007dr}, $430\,{\rm MeV}\text{--}4\,{\rm TeV}$: Fermi-LAT \cite{Fermi-LAT:2015kyq, Foster:2022nva}, $4\,{\rm TeV}\text{--} 10\,{\rm TeV}$: AMS-02 \cite{Giesen:2015ufa}). 
    The solid blue curve shows the constraints from $J\rightarrow \bar{q}q$ ($4\,{\rm GeV}\text{--}20\,{\rm GeV}$: Voyager 1 / AMS-02 \cite{Boudaud:2016mos,Ibarra:2013zia,Giesen:2015ufa}, $20\,{\rm GeV}\text{--} 10\,{\rm TeV}$: Fermi-LAT \cite{Cohen:2016uyg}). 
    The solid red curve shows the constraints from $J\rightarrow e^{+}e^{-}$ ($1\,{\rm MeV}\text{--}1.2\,{\rm MeV},7.1\,{\rm MeV}\text{--}15\,{\rm MeV}$: CMB\cite{Slatyer:2016qyl}, $1.2\,{\rm MeV}\text{--}7.1\,{\rm MeV}$: Lyman-$\alpha$ \cite{Liu:2020wqz}, $15\,{\rm MeV}\text{--} 0.7\,{\rm GeV}, \,10\,{\rm GeV}\text{--} 0.6\,{\rm TeV}$: Voyager 1/AMS-02 \cite{Boudaud:2016mos,Ibarra:2013zia}, $0.7\,{\rm GeV}\text{--} 10\,{\rm GeV},\, 0.6\,{\rm TeV}\text{--}10\,{\rm TeV}$: Fermi-LAT \cite{Cohen:2016uyg}). 
    The solid yellow curve shows the constraints from $J\rightarrow \mu^{+}\mu^{-}$ ($0.1\,{\rm GeV}\text{--}1\,{\rm GeV},\,10\,{\rm GeV}\text{--}0.5\,{\rm TeV}$: Voyager 1/AMS-02 \cite{Boudaud:2016mos,Ibarra:2013zia}, $1\,{\rm GeV}\text{--} 10\,{\rm GeV},\, 0.5\,{\rm TeV}\text{--}10\,{\rm TeV}$: Fermi-LAT \cite{Cohen:2016uyg}).
    The solid green curve shows the constraints from $J\rightarrow \tau^{+}\tau^{-}$ ($2\,{\rm GeV}\text{--}4\,{\rm GeV}$: Voyager 1 \cite{Boudaud:2016mos}, $4\,{\rm GeV}\text{--} 10\,{\rm TeV}$: Fermi-LAT \cite{Cohen:2016uyg}). 
    Dashed curves describe the expected sensitivities of the future experiment, CTA \cite{Pierre:2014tra}.
    } 
    \label{fig:result_other}
\end{figure}

We show the limits on the linear combinations of the components of $K$ defined in Eq.\ \eqref{Eq:K} from majoron visible decay channels $J\rightarrow \bar{f}f$ at one-loop and the value of $K'$ defined in Eq.\ \eqref{eq:Kprime} from $J\rightarrow \gamma\gamma$ at two-loop, using the latest data of gamma-ray and cosmic-ray experiments \cite{Fischer:2022pse,Yuksel:2007dr,Fermi-LAT:2012pls,Fermi-LAT:2015kyq,Cohen:2016uyg,Foster:2022nva,DiMauro:2023qat,MAGIC:2018tuz,Ninci:2019njk,Slatyer:2016qyl,Ibarra:2013zia,Boudaud:2016mos,Giesen:2015ufa,Liu:2020wqz} and analysis on the sensitivity of the future gamma-ray experiment, CTA \cite{Pierre:2014tra}.
The results are summarized in Figure.\ \ref{fig:result_other}. Purple, blue, red, yellow and green curves correspond to DM decays into $\gamma\gamma,\bar{q}q,e^{+}e^{-},\mu^{+}\mu^{-},\tau^{+}\tau^{-}$. Solid curves describe the current constraints, while dashed curves describe the expected sensitivities of the future experiment, CTA.

The decay channel $J\rightarrow \gamma\gamma$ constrains $K'$ defined in Eq.\ \eqref{eq:Kprime}. 
Gamma-ray telescopes have derived the upper limit on DM decay rate into photons in the mass range of $1\,{\rm MeV}\text{--}4\times 10^{3}\,{\rm GeV}$.
More concretely, INTEGRAL/SPI \cite{Fischer:2022pse}, and COMPTEL/EGRET \cite{Yuksel:2007dr} provide the strongest constraints in the DM mass range $1\,{\rm MeV}\leq m_{J}\leq 7\,{\rm MeV}$, and $7\,{\rm MeV}\leq m_{J}\leq 430\,{\rm MeV}$, respectively. 
Fermi-LAT gamma-ray search in the Milky-Way Halo and Galactic center \cite{Fermi-LAT:2015kyq, Foster:2022nva} provide the strongest constraints in the mass region of $430\,{\rm MeV}\text{--}4\,{\rm TeV}$, and anti-proton data by the cosmic-ray telescope AMS-02 \cite{Giesen:2015ufa} provides the constraints in higher energy region.
The combined result of those gamma-ray and anti-proton searches is shown in purple in Figure.\ \ref{fig:result_other}. 
We do not show the corresponding constraint from CMB \cite{Slatyer:2016qyl} since it is at least one order of magnitude weaker than the ones from those telescope observations. 
Note that even though the lifetime of the photon-inducing channel is severely constrained, the limitation on $K$ is at almost the same order or even weaker compared to that of fermion-inducing decay channels, because of the suppression by the factor $\alpha^{2}$ in Eq.\ \eqref{eq:decay rate into photons}.

The trace of $K$ is constrained from the decay channel $J\rightarrow \bar{q}q$.
Ref.\ \cite{Cohen:2016uyg} has analyzed Fermi-LAT data and derived the constraints on DM decays into $\bar{b}b$ and $\bar{t}t$ in the mass range of $m_{J}\gtrsim 20\,{\rm GeV}$ and $m_{J}\gtrsim 400\,{\rm GeV}$, respectively, which provide the strongest constraints on the trace of $K$ in $m_{J}\gtrsim 20 \,{\rm GeV}$. 
Cosmic positron searches by Voyager 1 \cite{Boudaud:2016mos}, and anti-proton searches by AMS-02 \cite{Giesen:2015ufa} also provide the constraints, which are strongest in the mass region of $4\text{--}20\,{\rm GeV}$.
The combined result of those gamma-ray and cosmic-ray searches is shown as a solid blue curve in Figure.\ \ref{fig:result_other}. 
Ref.\ \cite{Pierre:2014tra} has derived the expected sensitivity of CTA to DM decays into $\bar{b}b$, assuming 200 hours data-taking of Galactic center. 
We reinterpret it into the constraint on the trace of $K$, and show the result as a dashed blue curve in Figure.\ \ref{fig:result_other}. 
Since we consider the expected sensitivity of CTA to only the decay channel into $\bar{b}b$, not $\bar{t}t$,
the reinterpreted constraint based on Fermi-Lat analysis \cite{Cohen:2016uyg} is even much stronger than the expected sensitivity of CTA in the mass region $m\gtrsim 400\,{\rm GeV}$.
CMB search \cite{Slatyer:2016qyl,Garcia-Cely:2017oco}, cosmic positron observations by AMS-02 \cite{Ibarra:2013zia} and other gamma-ray searches by Fermi-LAT \cite{Fermi-LAT:2012pls,DiMauro:2023qat} and MAGIC \cite{MAGIC:2018tuz} provide the slightly weaker constraints, therefore not shown in Figure.\ \ref{fig:result_other}.

The constraints on the DM decay into charged leptons $\bar{l}l\ (l=e,\mu,\tau)$ are interpreted into the limits on the linear combination of the components of the $K$-matrix, $|-2K_{ll}+{\rm tr{K}}|$. In Figure.\ \ref{fig:result_other}, we show the results of $l=e,\mu,\tau$ cases in red, yellow and green, respectively. The solid curves corresponds to the current limit, and the dashed curves corresponds to the future sensitivities of CTA experiment \cite{Pierre:2014tra}.

For $J\rightarrow e^{+}e^{-}$, CMB observation \cite{Slatyer:2016qyl} and Lyman-$\alpha$ measurement \cite{Liu:2020wqz} provide the constraints in the widest mass range of $1\ {\rm MeV}$--$10\,{\rm TeV}$. The constraint from Lyman-$\alpha$ measurement is based on the model of intergalactic medium temperature, and we use the conservative result.
Cosmic positron observations by Voyager 1/AMS-02\cite{Boudaud:2016mos,Ibarra:2013zia} provide the strongest constraints in the mass range $15\,{\rm MeV}\text{--}0.7\,{\rm GeV}$, and $10\,{\rm GeV}\text{--}0.6\,{\rm TeV}$. Gamma-ray observation by Fermi-Lat \cite{Cohen:2016uyg} provides the strongest constraint in the other range of $m_{J}$. The combined result is shown in red in Figure.\ \ref{fig:result_other}.

For $J\rightarrow \mu^{+}\mu^{-}$, cosmic positron observations by Voyager 1/AMS-02\cite{Boudaud:2016mos,Ibarra:2013zia} provide the strongest constraints in the mass range $0.1\,{\rm GeV}\text{--}1\,{\rm GeV}$, and $10\,{\rm GeV}\text{--}0.5\,{\rm TeV}$. Gamma-ray observation by Fermi-Lat \cite{Cohen:2016uyg} provides the strongest constraint in the other range of $m_{J}$. The combined result is shown in yellow in Figure.\ \ref{fig:result_other}.
Ref.\ \cite{Pierre:2014tra} has derived the expected sensitivity of CTA to DM decays into $\mu^{+}\mu^{-}$, assuming 200 hours data-taking of Galactic center. The corresponding result is shown as a yellow dashed curve in Figure.\ \ref{fig:result_other}. The Fermi-Lat analysis \cite{Cohen:2016uyg} and Voyager 1/AMS-02\cite{Boudaud:2016mos,Ibarra:2013zia} provide slightly stronger constraints than the expected sensitivity of CTA. This might be because the existing observations deal with the larger data obtained from both galactic and extragalactic observations.  
CMB search \cite{Slatyer:2016qyl,Garcia-Cely:2017oco}, gamma-ray search by MAGIC \cite{MAGIC:2018tuz} and cosmic anti-proton search by AMS-02 \cite{Giesen:2015ufa} provide the slightly weaker constraints, therefore not shown in Figure.\ \ref{fig:result_other}.

For $J\rightarrow \tau^{+}\tau^{-}$, cosmic positron observations by Voyager 1 \cite{Boudaud:2016mos} and gamma-ray observations by Fermi-LAT \cite{Cohen:2016uyg} provide the strongest constraints in the mass range $2\ {\rm GeV}\text{--}4\,{\rm GeV}$, $4\ {\rm GeV}\text{--}10\,{\rm TeV}$, respectively. The combined result is shown as a solid green in Figure.\ \ref{fig:result_other}.
CMB search \cite{Slatyer:2016qyl,Garcia-Cely:2017oco}, cosmic positron observations by AMS-02 \cite{Ibarra:2013zia} and other gamma-ray searches by Fermi-LAT\cite{Fermi-LAT:2012pls,DiMauro:2023qat} and MAGIC \cite{MAGIC:2018tuz} provide the slightly weaker constraints, therefore not shown in Figure.\ \ref{fig:result_other}.
Ref.\ \cite{Pierre:2014tra} has derived the expected sensitivity of CTA to DM decays into $\tau^{+}\tau^{-}$, assuming 200 hours data-taking of Galactic center. 
The reinterpreted constraint is shown as a dashed green curve in Figure.\ \ref{fig:result_other}.
The Fermi-Lat analysis \cite{Cohen:2016uyg} provides slightly stronger constraints than the expected sensitivity of CTA. This might be because the existing observations deal with the larger data obtained from both galactic and extragalactic observations.

\section{Summary and conclusions}\label{sec:conclusions}

In this work, we have updated constraints and discussed future prospects on majoron DM in the singlet majoron model in the mass region of MeV--10 TeV from neutrino telescopes as in Figures.~\ref{fig:figure NH} and \ref{fig:figure QD}, and gamma-ray and cosmic-ray telescopes as in Figure.~\ref{fig:result_other}. The detail of the update from the privious work by Garcia-Cely and Heeck \cite{Garcia-Cely:2017oco} is discussed in appendix~\ref{app:comparison}.

From neutrino signals, we can constrain the lepton number breaking scale, $f$, in the singlet majoron model.
In Figures.~\ref{fig:figure NH} and \ref{fig:figure QD}, we show the constraints on $f$ in the NH and QD cases, respectively. We do not show the IH case since the constraints in the IH case appear in between the NH and QD cases. This is because the decay rates are proportional to neutrino mass squared and in the NH case, $\nu_e$-flux is suppressed by small mixing between electron-flavor and mass-eigenstates of the heaviest active neutrino.
We have extended the mass range up to $10\,{\rm TeV}$ and updated the current constraints from Borexino, KamLAND, SK, and added new current constraints from IceCube and ANTARES and expected constraints from JUNO, HK, and P-ONE. 
Those figures show that future neutrino detectors will update the current constraint within (2-3)-fold in the mass region $(5-100)\,{\rm MeV}$, and nearly one order of magnitude or even more in the mass region $(0.5-10)\,{\rm GeV}$ in any mass hierarchy cases. 
DUNE and KM3NeT will also improve the constraints on majoron DM significantly, but the contribution of each flavor neutrinos to these searches are non-trivial. We leave analyses of DUNE and KM3NeT as future work.

From other visible signals, we can constrain $K$ and $K'$ defined by Eq.~(\ref{Eq:K})( or Eq.~(\ref{Eq:K2})) and Eq.\ \eqref{eq:Kprime}, respectively, which are the combination of the lepton number breaking scale and the other seesaw parameters in Eq.~(\ref{Lsingletmajoron}).
The CMB, gamma-ray and cosmic-ray observations provide complementary constraints from neutrino signal. We have extended the mass range up to $10\,{\rm TeV}$, updating the current constraints from INTEGRAL/SPI, Fermi-LAT, AMS-02, and adding the current constraints by Voyager 1 and MAGIC and the future sensitivity by CTA as in Figure.\ \ref{fig:result_other}.

\section*{Acknowledgments}
We would like to thank Yoshihiko Abe and Yu Hamada for valuable discussions. We also thank Julian Heeck, Juntaro Wada, Keiichi Watanabe, Masahide Yamaguchi and the anonymous referee for useful comments. KA is supported by IBS under the project code, IBS-R018-D1. MN is supported by IBS under the project code, IBS-R018-D3.

%%%%%%%%%%%%%%%%%%%%%%%%%%%%%%%%%%%%%%%%%%%%%%%%%%%%%%%%%%%%%%%%%%%%%%%%%%%%%%%%%%%%%%%%%%%%%%%%%%
%%%%%%%%%%%%%%%%%%%%%%%%%%%%%%%%%%%%%%%%%%%%%%%%%%%%%%%%%%%%%%%%%%%%%%%%%%%%%%%%%%%%%%%%%%%%%%%%%%
%%%%%%%%%%%%%%%%%%%%%%%%%%%%%%%%%%%%%%%%%%%%%%%%%%%%%%%%%%%%%%%%%%%%%%%%%%%%%%%%%%%%%%%%%%%%%%%%%%

\appendix
\section{Comparison with the literature}\label{app:comparison}

In this appendix, we discuss how the constraints on majoron DM in the singlet majoron model is updated compared with the previous work by Garcia-Cely and Heeck \cite{Garcia-Cely:2017oco}. The references to data and prior analyse used to produce the results in this paper are listed in Tables.\ \ref{tab:data_neutrino} and \ref{tab:data_other}.
First, we extend the previous constraints in the range of majoron mass of MeV--100 GeV to MeV--10 TeV. 

For the decay channel into neutrinos, the constraints on $f$ from the data of Borexino in 2010 \cite{Borexino:2010zht}, KamLAND in 2011 \cite{KamLAND:2011bnd}, DSNB search in SK-I \cite{Super-Kamiokande:2002exp}, in SK-IV within 960 days \cite{Super-Kamiokande:2013ufi} and atmospheric neutrino search in SK \cite{Palomares-Ruiz:2007egs,Frankiewicz:2016nyr} are used in Figure.\ 4 in Ref.\ \cite{Garcia-Cely:2017oco}. 
After nearly 10 years of additional data acquisition in the Borexino and KamLAND experiments, we show that the $f$ constraints can be updated to within (2-3)-fold. The constraint on $f$ from DSNB search in SK is updated within nearly two-hold, using SK-IV data during 2970 days \cite{Super-Kamiokande:2021jaq}, shown in red in Figures.\ \ref{fig:figure NH}, and \ref{fig:figure QD}. 
We use the same data and analysis as Ref.\ \cite{Garcia-Cely:2017oco} for atmospheric neutrino search, shown in light-blue in Figure.\ \ref{fig:figure NH}, and \ref{fig:figure QD}.
Furthermore, we add the constraints from IceCube \cite{IceCube:2021kuw,IceCube:2023ies} and ANTARES \cite{Albert:2016emp,Arguelles:2022nbl} experiments, and the expected sensitivities of JUNO \cite{Akita:2022lit}, HK \cite{Bell:2020rkw}, and P-ONE \cite{P-ONE:2020ljt,Arguelles:2022nbl}.
They also use the cosmological constraint on the DM model-independent lifetime $\Gamma < (5\times 10^8\,{\rm sec})^{-1}$ \cite{Audren:2014bca} in Ref.\ \cite{Garcia-Cely:2017oco}, and we updated it with $\Gamma < (250\,{\rm Gyr})^{-1} \sim (8\times 10^8\,{\rm sec})^{-1}$ \cite{Alvi:2022aam, Simon:2022ftd}, which is shown in black in Figures.\ \ref{fig:figure NH}, and \ref{fig:figure QD}.

For the decay channel into photons, the constraints on $K$ from the data of INTEGRAL/SPI in 2007 \cite{Boyarsky:2007ge}, COMPTEL/EGRET in 2007 \cite{Yuksel:2007dr}, and Fermi-LAT in 2015 \cite{Fermi-LAT:2015kyq} are shown in purple in Figure.\ 5 in Ref.\ \cite{Garcia-Cely:2017oco}, neglecting the contribution from $W$-boson.
We use the same result of COMPTEL/EGRET and updated other constraints using the recent analysis based on the data by INTEGRAL/SPI in 2022 \cite{Fischer:2022pse} and Fermi-LAT in 2015 and 2022 \cite{Fermi-LAT:2015kyq,Foster:2022nva}, and derive the constraint on $K'$ defined in Eq.\ \eqref{eq:Kprime} taking the $W$-boson contribution into account. In the mass range $m_{J}>4\times 10^{6}\,{\rm MeV}$, the current strongest constraint comes from AMS-02 anti-proton search \cite{Giesen:2015ufa}. The combined result is shown in purple in Figure.\ \ref{fig:result_other}.

For the decay channel into bottom quarks, the constraints from Fermi-LAT in 2016 \cite{Cohen:2016uyg}, CMB \cite{Slatyer:2016qyl} and anti-proton flux observation by AMS-02 in 2015 \cite{Giesen:2015ufa} is shown in Figure.\ 5 of \cite{Garcia-Cely:2017oco}. Even though several gamma-ray telescope experiments \cite{DiMauro:2023qat,MAGIC:2018tuz,Ninci:2019njk} put the constraints on this DM decay channel, the current strongest constraints are obtained from Fermi-LAT in 2016 \cite{Cohen:2016uyg}.
We also newly show the expected sensitivity of CTA experiment \cite{Pierre:2014tra} as a blue dashed curve.

In Ref.\ \cite{Garcia-Cely:2017oco}, the constraints from electron/positron flux observations by AMS-02 \cite{Ibarra:2013zia} are presented as dotted curves in red, black and green in the mass range $(10-100)\,{\rm GeV}$, which correspond to the decay channels $J\rightarrow e^+e^-,\mu^+\mu^-,\,{\rm and}\,\tau^+\tau^-$, respectively. 
We extend the mass region to constrain by adding the result obtained in Ref.\ \cite{Boudaud:2016mos, Cohen:2016uyg, Liu:2020wqz}, as shown in red, yellow and green in Figure.\ \ref{fig:result_other}, respectively.
We also show the expected sensitivity of CTA \cite{Pierre:2014tra} experiment for $J\rightarrow \mu^+\mu^-$ and $J\rightarrow \tau^+\tau^-$ as yellow and green dashed curves, respectively.

\begin{table}[]
    \centering
    \begin{tabular}{l|lll}
    Experiments & Mass range & Flavors & Data and prior analysis used in this paper\\ \hline
    Borexino &$(3.6\text{--}21.6) \,{\rm MeV}$&$e$&Upper limit on DSNB flux\cite{Borexino:2019wln}\\
    KamLAND &$(16.6\text{--}61.6)\, {\rm MeV}$&$e$&Upper limit on DSNB flux\cite{KamLAND:2021gvi}\\
    SK, DSNB &$(18.6\text{--}62.6)\, {\rm MeV}$&$e$&Upper limit on DSNB flux\cite{Super-Kamiokande:2011lwo,Super-Kamiokande:2021jaq}\\
    SK &$(20\text{--}400)\, {\rm MeV}$&$e$&Annihilation Cross Section \cite{Olivares-DelCampo:2017feq}\\
    SK Atmospheric $\nu$ &$(0.1\text{--}10^{4})\, {\rm GeV}$&$\mu$&Lifetime \cite{Palomares-Ruiz:2007egs,Frankiewicz:2016nyr}\\
    IceCube Deep-Core&$(10\text{--}4\times 10^{2})\, {\rm GeV}$&all flavors&Annihilation Cross Section \cite{IceCube:2021kuw}\\
    IceCube&$(16\text{--}10^{4})\, {\rm GeV}$&all flavors&Lifetime \cite{IceCube:2023ies}\\
    ANTARES&$(16\text{--}10^{4})\, {\rm GeV}$&$\mu$&Lifetime \cite{Albert:2016emp,Arguelles:2022nbl}\\
    JUNO&$(5\text{--}200)\, {\rm MeV}$&$e$&Lifetime \cite{Akita:2022lit}\\
    HK&$(0.03\text{--}100)\,{\rm GeV}$&$e,\mu$&Annihilation Cross Section\cite{Bell:2020rkw}\\
    P-ONE&$(\mathcal{O}(1)\text{--}10)\,{\rm TeV}$&$\mu$&Lifetime\cite{P-ONE:2020ljt,Arguelles:2022nbl}
    \end{tabular}
    \caption{List of data sets and prior analysis used in this work to produce Figures.\ \ref{fig:figure NH}, and \ref{fig:figure QD}.}
    \label{tab:data_neutrino}
\end{table}

\begin{table}[]
    \centering
    \begin{tabular}{l|ll}
    Decay channel & Experiments and Prior analysis &Mass Region\\ \hline
    $J\rightarrow \gamma\gamma$&
    INTEGRAL/SPI\cite{Fischer:2022pse} 
    &$(1\times 10^{-3}\text{--}7\times 10^{-3})\,{\rm GeV}$\\
    &COMPTEL/EGRET\cite{Yuksel:2007dr}
    &$(1\times 10^{-3}\text{--}2\times 10^{2})\,{\rm GeV}$\\
    &Fermi-LAT\cite{Fermi-LAT:2015kyq}
    &$(4\times 10^{-1}\text{--}1\times 10^{3})\,{\rm GeV}$\\
    &Fermi-LAT\cite{Foster:2022nva}
    &$(2\times 10^{1}\text{--}4\times 10^{3})\,{\rm GeV}$\\

    &CMB\cite{Slatyer:2016qyl}    &$(1\times 10^{-3}\text{--}1\times 10^{4})\,{\rm GeV}$\\

    &AMS-02 ($\bar{p}$)\cite{Giesen:2015ufa}    
    &$(1\times 10^{1}\text{--}1\times 10^{4})\,{\rm GeV}$\\
    \hline
    $J\rightarrow \bar{b}b$
    &Fermi-LAT \cite{Fermi-LAT:2012pls}    
    &$(1\times 10^{1}\text{--}1\times 10^{4})\,{\rm GeV}$\\

    &Fermi-LAT\cite{Cohen:2016uyg}    &$(2\times 10^{1}\text{--}1\times 10^{4})\,{\rm GeV}$\\

    &Fermi-LAT\cite{DiMauro:2023qat}    &$(5\times 10^{0}\text{--}1\times 10^{4})\,{\rm GeV}$\\

    &MAGIC \cite{MAGIC:2018tuz}
    &$(2\times 10^{2}\text{--}1\times 10^{4})\,{\rm GeV}$\\
    
    &MAGIC \cite{Ninci:2019njk}
    &$(3\times 10^{2}\text{--}1\times 10^{4})\,{\rm GeV}$\\

    &CTA\cite{Pierre:2014tra}
    &$(7\times 10^{2}\text{--}1\times 10^{4})\,{\rm GeV}$\\
    &CMB\cite{Slatyer:2016qyl}
    &$(1\times 10^{1}\text{--}1\times 10^{4})\,{\rm GeV}$\\

    &Voyager 1/AMS-02($e^{+}$) \cite{Ibarra:2013zia,Boudaud:2016mos}
    &$(4\times 10^{0}\text{--}2\times 10^{3})\,{\rm GeV}$\\

    &AMS-02($\bar{p}$) \cite{Giesen:2015ufa}
    &$(1\times10^{1}\text{--}1\times 10^{4})\,{\rm GeV}$\\
    \hline
    $J\rightarrow \bar{t}t$&Fermi-LAT\cite{Cohen:2016uyg}&$(4\times 10^{2}\text{--}1\times 10^{4})\,{\rm GeV}$\\\hline
    $J\rightarrow e^{+}e^{-}$&Fermi-LAT\cite{Cohen:2016uyg} &$(7\times 10^{-1}\text{--}1\times 10^{4})\,{\rm GeV}$\\
    &CMB\cite{Slatyer:2016qyl}&$(1\times 10^{-3}\text{--}1\times 10^{4})\,{\rm GeV}$\\
    &Lyman-$\alpha$\cite{Liu:2020wqz}&$(1\times 10^{-3}\text{--}1\times 10^{3})\,{\rm GeV}$\\
    &Voyager 1/ AMS-02($e^{+}$) \cite{Ibarra:2013zia,Boudaud:2016mos}&$(1\times 10^{-2}\text{--}2\times 10^{3})\,{\rm GeV}$\\  \hline
    $J\rightarrow \mu^{+}\mu^{-}$
    &Fermi-LAT\cite{Fermi-LAT:2012pls}
    &$(1\times 10^{1}\text{--}1\times 10^{4})\,{\rm GeV}$\\
    &Fermi-LAT\cite{Cohen:2016uyg}&$(7\times 10^{-1}\text{--}1\times 10^{4})\,{\rm GeV}$\\
    &MAGIC\cite{MAGIC:2018tuz}
    &$(2\times 10^{2}\text{--}1\times 10^{4})\,{\rm GeV}$\\
    &CTA\cite{Pierre:2014tra}
    &$(7\times 10^{2}\text{--}1\times 10^{4})\,{\rm GeV}$\\
    &CMB\cite{Slatyer:2016qyl}&$(1\times 10^{1}\text{--}1\times 10^{4})\,{\rm GeV}$\\
    &Voyager 1/AMS-02($e^{+}$) \cite{Ibarra:2013zia,Boudaud:2016mos}
    &$(1\times 10^{-1}\text{--}2\times 10^{3})\,{\rm GeV}$\\
    &AMS-02($\bar{p}$) \cite{Giesen:2015ufa}
    &$(3\times 10^{2}\text{--}1\times 10^{4})\,{\rm GeV}$ \\ \hline
    $J\rightarrow \tau^{+}\tau^{-}$
    &Fermi-LAT\cite{Fermi-LAT:2012pls}&$(1\times 10^{1}\text{--}1\times 10^{4})\,{\rm GeV}$ \\ 
    &Fermi-LAT\cite{Cohen:2016uyg}&$(4\times 10^{0}\text{--}1\times 10^{4})\,{\rm GeV}$\\
    &Fermi-LAT\cite{DiMauro:2023qat}&$(5\times 10^{0}\text{--}1\times 10^{4})\,{\rm GeV}$ \\ 
    &MAGIC\cite{MAGIC:2018tuz}&$(2\times 10^{2}\text{--}1\times 10^{4})\,{\rm GeV}$ \\ 
    &CTA\cite{Pierre:2014tra}
    &$(7\times 10^{2}\text{--}1\times 10^{4})\,{\rm GeV}$\\
    &CMB\cite{Slatyer:2016qyl}&$(1\times 10^{1}\text{--}1\times 10^{4})\,{\rm GeV}$ \\ 
    &Voyager 1/AMS-02($e^{+}$) \cite{Ibarra:2013zia,Boudaud:2016mos}&$(2\times 10^{0}\text{--}2\times 10^{3})\,{\rm GeV}$ \\ 
    \end{tabular}
    \caption{List of data sets and prior analysis used in this work to produce Figure.\ \ref{fig:result_other}.}
    \label{tab:data_other}
\end{table}

\bibliography{reference} 
\bibliographystyle{JHEP}

\end{document}